# Comparison of Electroluminescence and Photoluminescence Imaging of Mixed-Cation Mixed-Halide Perovskite Solar Cells at Low Temperatures

*Hurriyet Yuce-Cakir,[1,2]\* Haoran Chen,[3] Isaac Ogunniranye,[3] Susanna M. Thon,[2,4,5] Yanfa Yan,[3] Zhaoning Song,[3] and Behrang H. Hamadani[6]\**

[1]PREP Associate, Engineering Laboratory, National Institute of Standards & Technology, Gaithersburg, Maryland 20899, United States.
[2]Whiting School of Engineering, Ralph O`Connor Sustainable Energy Institute, Johns Hopkins University, Baltimore, Maryland 21218, United States.
[3]Department of Physics and Astronomy and Wright Center for Photovoltaics Innovation and Commercialization, The University of Toledo, Toledo, Ohio 43606, United States.
[4]Department of Electrical and Computer Engineering, Johns Hopkins University, Baltimore, Maryland 21218, United States.
[5]Department of Materials Science and Engineering, Johns Hopkins University, Baltimore, Maryland 21218, United States.
[6]Engineering Laboratory, National Institute of Standards & Technology, Gaithersburg, Maryland 20899, United States.

E-mail: hurriyet.yucecakir@nist.gov, behrang.hamadani@nist.gov

Keywords: perovskite solar cell, hyperspectral imaging, electroluminescence, photoluminescence, charge injection

**Abstract**

Halide perovskites have emerged as promising candidates for high-performance solar cells. This study investigates the temperature-dependent optoelectronic properties of mixed-cation mixed-halide perovskite solar cells using electroluminescence (EL) and photoluminescence (PL) hyperspectral imaging, along with current–voltage analysis. Luminescence images, which were converted to EL and PL external radiative efficiency (ERE) maps, revealed significant changes in the optoelectronic behavior of these devices at low temperatures. Specifically, we found that a significant source of heterogeneity in the low-temperature EL ERE maps below 240 K is related to local charge injection and extraction bottlenecks, whereas PL ERE maps show suppressed non-radiative recombination and significant improvements in efficiency throughout the investigated temperature range. The spatial distribution of ERE and its variation with applied current were analyzed, offering insights into charge-carrier dynamics and defect behavior. Our results reveal that while the perovskite layer exhibits enhanced ERE at low temperatures, charge injection barriers at the interfaces of the perovskite solar cells significantly suppress EL and degrade the fill factor below 240 K. These findings reveal that a deeper understanding of the performance of perovskite solar cells under low-temperature conditions is an essential step toward their potential application in space power systems and advanced semiconductor devices.

1. Introduction

Metal halide perovskites have emerged as promising energy materials for high-performance solar cells due to their exceptional optical and electrical properties, including high absorption coefficients, low trap densities, high carrier mobilities, and long carrier diffusion lengths.[1,2] Since 2009, the power conversion efficiency (PCE) of perovskite solar cells has risen dramatically, now exceeding 26 %.[3] This remarkable improvement, combined with low-cost fabrication processes, makes perovskite solar cells a highly attractive photovoltaic technology.[4,5]

As research continues to advance the performance and stability of perovskite solar cells (PSCs), the use of advanced characterization techniques becomes increasingly important to investigate the underlying physical mechanisms linked to their optoelectronic properties.[5–7] Among these techniques, electroluminescence (EL) and photoluminescence (PL) imaging provide valuable insights into charge carrier recombination dynamics, defect distribution, and material inhomogeneities.[8–13] However, EL and PL measurements at room or elevated temperatures may not provide a complete picture of recombination phenomena or charge transport within the semiconductor active layer because some processes, such as certain radiative transitions,[14] phase changes,[15] and charge injection through contacts,[16] are temperature dependent. Therefore, low-temperature luminescence measurements are often preferred for a more detailed examination of defect-related and other physical properties, which directly impact recombination mechanisms and device performance.[17–19] Low temperature characterization is also important for deployment of PSCs in space applications where they are subjected to temperature cycling over extended periods including temperatures as low as 170 K.[20–22]

Upon optical or electrical stimulation, excess charge carriers can be trapped at defect sites arising from lattice dislocations, grain boundaries, and/or point defects, leading to both radiative and non-radiative recombination pathways.[23,24] By analyzing both the emission peak energetics and their intensity as a function of temperature, critical insights can be gained into the nature of existing defects in the material. Comparing EL and PL imaging at low temperatures offers a comprehensive understanding of radiative versus non-radiative recombination mechanisms, interface quality, and local non-uniformities within the perovskite layer. Despite a growing number of studies investigating low-temperature behavior in perovskites[18,25,26], low-temperature EL and its comparison with PL imaging of the mixed-cation and mixed-anion perovskites remain relatively unexplored.

In this work, we report on the low-temperature optical and electrical behavior of $Rb_{0.05}Cs_{0.05}MA_{0.05}FA_{0.85}Pb(I_{0.95}Br_{0.05})_3$ perovskite solar cells. In a novel approach, we performed both absolute EL and PL hyperspectral imaging across the entire device area as a function of temperature, complemented by dark and light $J$–$V$ analyses from 300 K to 160 K. EL and PL absolute photon flux maps were converted to external radiative efficiency (ERE) maps to visualize changes within the active layer more effectively at low temperatures. As the temperature decreased, the mean EL performance initially improved down to ≈ 240 K, then declined, which we attributed to limitations in charge injection through the contacts. In contrast, PL showed a continuous, systematic increase in radiative efficiency. Furthermore, the differences in the EL and PL ERE maps at all temperatures indicate that a combination of local contact effects and local nonradiative recombination inhomogeneities, such as local charge trapping or defect sites, affect luminescence mapping in these materials. The electrical characterization results are strongly in agreement with the luminescence imaging data. These findings shed light on both the potential and challenges of operating perovskite solar cells at low temperatures. Understanding charge carrier dynamics and optoelectronic behavior under these conditions is critical for advancing perovskite solar cell applications, particularly in power generation for space environments.[27]

2. Results and Discussion

We fabricated inverted p-i-n perovskite solar cells with the device architecture FTO/MeO-2PACz/perovskite/$C_{60}$/BCP/Ag following a reported procedure (for details about the device, see Experimental section).[28] RbCsMAFA-based perovskite solar cells were optically and electrically characterized at low temperatures in an optical cryostat using liquid nitrogen. The mean absolute EL spectral photon flux curves under a 14 mA/cm$^2$ injection current, measured across the entire cell area, are shown in **Figure 1**a for temperatures ranging from 300 K to 160 K. As the device temperature decreases from 300 K to 240 K, the photon flux systematically increases due to a reduction in nonradiative recombination rates. However, further cooling to 160 K results in a decrease in the mean emitted photon flux, eventually falling below the initial intensity observed at 300 K. In contrast, the PL photon flux, which was measured by exciting the perovskite uniformly with a 532-nm laser over the same device and collecting the PL signal, shows a systematic increase as the temperature decreases from 300 K to 120 K, as shown in **Figure 1**b. The increase in PL intensity with decreasing temperatures is consistent with what has been observed in traditional semiconductors.[29,30] At low temperatures, deep trap states are effectively frozen out, suppressing trap-assisted nonradiative recombination and resulting in higher luminescence.[31] Additionally, a phase transition from cubic to tetragonal at low temperatures has been reported to facilitate the removal of intrinsic point defects in the perovskite absorber layer, further enhancing luminescence.[32]

Considering the gradual enhancement of PL flux intensity with decreasing temperature, one strong possibility for the decline in EL photon flux intensity below 240 K is a reduction in charge carrier density within the perovskite absorber layer. Since EL is performed in the dark, its strength depends on the radiative recombination of injected electrons and holes from the contacts into the perovskite layer. If carrier injection into the active layer is impeded or suppressed at the ETL/PSC or HTL/PSC interfaces due to the formation of an energetic barrier, then we can expect to see a reduction in the EL intensity. **Figure 1**c provides a schematic illustration of this charge injection issue between the perovskite layer and the charge transport layers during EL measurements at low temperatures. The device structure plays an important role in charge injection at low temperatures. It should be noted that an increase in EL intensity with decreasing temperature has been reported for perovskite light-emitting devices that incorporate discrete perovskite platelets embedded in an organic insulator.[33] Therefore, our different observation here is likely related to charge injection issues and will be further investigated below.

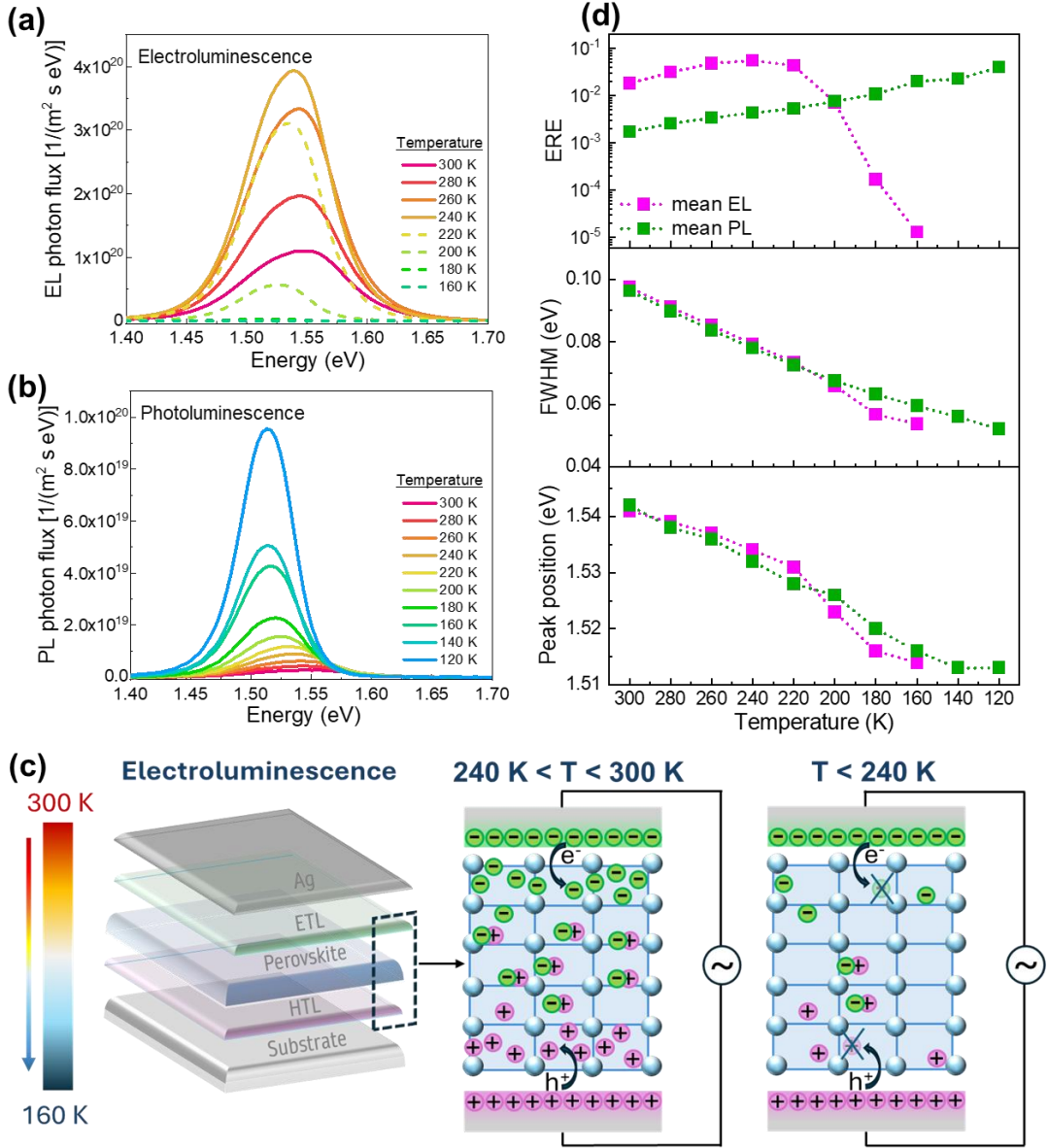

**Figure 1.** (a) The EL and (b) the PL emission spectra of the perovskite solar cell at a range of temperatures. (c) Schematic illustration of the hindered charge injection in perovskite solar cells at low temperatures (right) compared to higher temperatures (middle). (d) Comparison of the temperature-dependent ERE, the FWHM, and peak positions of the EL and PL spectrum of the perovskite solar cells.

We converted the EL and PL photon fluxes to EL ERE and PL ERE, respectively, using the relationship: EL ERE $= \frac{q\, \varphi_{ext}^{tot}}{J_{inj}}$ and PL ERE $= \frac{\varphi_{ext}^{tot}}{\varphi_{abs}}$ where $q$ is the electron charge, $\varphi_{ext}^{tot}$ is the total external luminescence photon flux in photons/(m²·s), $J_{inj}$ is the injection current density, and $\varphi_{abs}$ is the absorbed excitation photon flux. ERE is a very powerful parameter to assess the voltage losses from the internal voltage and is directly related to nonradiative recombination losses within the device stack. An ERE of unity indicates that for every injected electron into the device (EL) or every absorbed incident excitation photon (PL), one photon is externally emitted by the device, indicating 100 % external radiative efficiency. In photovoltaic devices, a cell with an ERE value in the range ≈ 0.01 to 0.1 is considered excellent.[34] ERE can also have a substantial dependence on the incident photon flux or injection current magnitudes, and, therefore, the EL ERE and the PL ERE may not necessarily be equal. **Figure 1**d (top panel) shows that the calculated mean EL and PL ERE behaviors with temperature are consistent with their respective trends in photon flux (**Figure 1**a,b). Yet, reporting the efficiency of the luminescence process in terms of ERE is a more meaningful approach because it is tied to the fundamental physics of solar cells. The observed higher EL ERE compared to PL ERE at 300 K is due to differences in experimental conditions: the injection current was optimized to obtain clear EL images even at low temperatures, while the laser power for PL measurements was kept low to avoid sample heating and degradation of the perovskite material during testing and particularly at low temperatures. For the PL measurements, the incident photon flux is ≈ $2.5 \times 10^{16}$ photons/(m$^2$.s) whereas for the EL measurements, the electron injection flux is ≈ $8.8 \times 10^{16}$ electrons/(m$^2$.s). In this case, EL ERE is expected to be higher than PL ERE since ERE is injection dependent.

**Figure S1**a, b (Supporting Information) show the normalized EL and PL spectra of the perovskite device across the same temperature range. Both EL and PL peak positions, centered at 1.54 eV at 300 K, gradually shift to 1.51 eV as the temperature decreases, accompanied by a narrowing of the emission peak as measured by the full-width-at-half-maximum (FWHM) of the signal, shown in **Figure 1**d. The redshift of the PL peaks with decreasing temperature originates from the shrinkage of the perovskite lattice, which enhances the overlap between Pb-6s and I-5p antibonding atomic orbitals that form the valence band maximum.[35] This behavior effectively causes a reduction in band gap energy as temperature is lowered. The reduced FWHM behavior with temperature is consistent with reduced electron–phonon coupling at lower temperatures.[36]

**Figure 2** shows the temperature evolution of an EL and PL ERE map collected from the same section of a perovskite cell between 300 K and 160 K. In **Figure 2**a, the EL ERE maps are derived from absolute hyperspectral EL images, representing the emitted photon flux in response to a 1 mA injection current. The EL ERE map at 300 K appears blotchy and nonuniform, with a large number of pinholes or patchy regions (blue regions) exhibiting lower ERE. As the temperature decreases below 220 K, the blue low-ERE patches grow larger, and eventually dominate the entire device area with only a few small hot spots (red areas) remaining at 160 K. In a dramatic contrast, the PL ERE distribution at 300 K in **Figure 2**b is quite uniform. At lower temperatures, the PL ERE map also shows some nonuniformity with a distribution that is different than that of the EL ERE map, but the magnitude of nonuniformity is significantly less than EL ERE as seen in the colorbar scale. Notably, the blue regions at low temperatures still show higher ERE than the red regions at higher temperatures, as all images are displayed using auto-scaling.

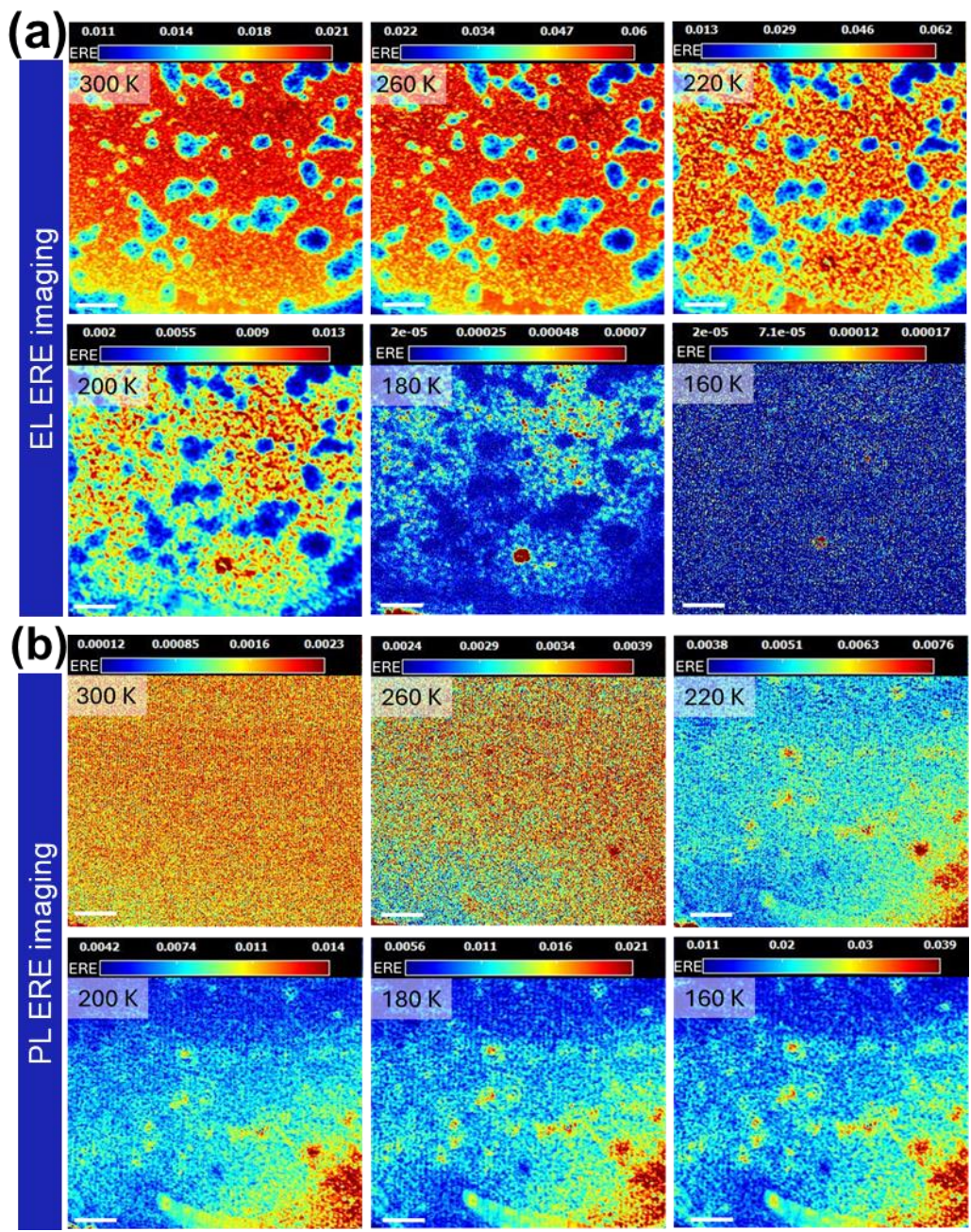

**Figure 2.** (a) EL and (b) PL ERE mapping over the same area of a perovskite solar cell at temperatures ranging from 300 K to 160 K. Scale bar is 200 µm.

In both EL and PL, photons are emitted from electron-hole radiative recombination events as a result of either excitation by injected charge carriers from the contacts (EL) or by incident photons (PL). Therefore, if a PL map shows a homogeneous region while the EL map of the same area shows significant inhomogeneity, that could indicate local regions where charge injection from the contacts is bottlenecked. The temperature dependent EL maps show that initially at room temperature, these bottlenecked regions are small and relatively efficient in bringing charge into the active layer. However, at temperatures below 200 K, these regions expand, and the overall efficiency of charge injection, even in brighter colored regions, decreases significantly, resulting in the nose-dive of the EL ERE as shown in **Figure 1**d.

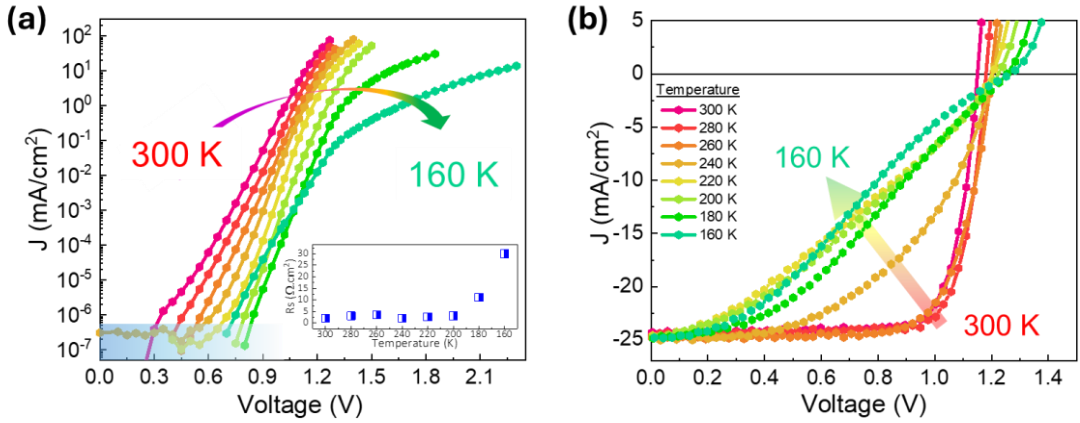

**Figure 3**. (a) Dark and (b) light J-V plots of the perovskite solar cell as a function of temperature. The inset in (a) shows series resistance extracted from the dark J-V as a function of temperature.

In order to gain better insight into the temperature dependent luminescence behavior of these devices, we also performed a series of temperature dependent dark and light *J-V* characterization measurements on several nominally-similar solar cells. **Figure 3**a shows the dark *J-V* plots of a perovskite cell measured between 300 K and 160 K. As the temperature decreases, the slope of the *J-V* curves at higher currents becomes shallower, indicating an increase in series resistance ($R_s$). The $R_s$ values have been extracted from the dark *J-V* data, and rise steadily with decreasing temperature, from 2.09 $\Omega \cdot cm^2$ at 300 K to 29.92 $\Omega \cdot cm^2$ at 160 K as shown in the inset of **Figure 3**a. This increase may be related to the existence of an energetic barrier such as a Schottky barrier between the perovskite and one or both of the electron or hole transport layers.[37] At low temperatures, it becomes energetically more difficult to inject electrons or holes from the contacts into the PSC layer over this energy barrier, and hence the series resistance of the device increases.

**Figure 3**b displays the light *J-V* measurements of the device under reverse scan over the same temperatures (see **Figure S2** and **S3**, Supporting Information for both reverse and forward scans, and external quantum efficiency spectrum, EQE). The device parameters extracted from the light *J-V* curves are summarized in **Figure S4** (Supporting Information). At 300 K, the perovskite solar cell exhibits a PCE of 22.08 %, with an open-circuit voltage ($V_{oc}$) of 1.15 V, a short-circuit current density ($J_{sc}$) of 24.28 mA/cm², and a fill factor (FF) of 78.7 %. As the temperature decreases from 300 K to 160 K, the $V_{oc}$ gradually increases from 1.15 V to 1.27 V, while the FF significantly drops from 78.7 % to 31.1 %. The $J_{sc}$ remains nearly constant at around 24.5 mA/cm² over this temperature range. The increase in $V_{oc}$ at lower temperatures can be explained by equation 1. Unlike traditional semiconductors, the bandgap energy ($E_g$) of perovskites decreases with decreasing temperatures, which would have a reducing effect on the $V_{oc}$. However, the net effect of the temperature reduction still favors an increase in the $V_{oc}$ much like traditional semiconductors because the middle term in equation 1, $\Sigma V_{loss}$, is a loss term related to the sum of all fixed entropic penalties that are temperature dependent, and the impact of these voltage penalties gets smaller with temperature.[38] The last term is also a loss term associated with the radiative efficiency of the cell. Although it can be substantial at room temperature (≈ 160 meV), its impact is reduced at low temperatures due to the $kT/q$ prefactor. In this equation, the PL ERE is the parameter that should be used in the calculation because it represents the intrinsic radiative efficiency of the perovskite material without the influence of the contacts. The PL ERE consistently increases as T is lowered, due to defect suppression at low temperatures.[39] This has a slightly beneficial effect for the $V_{oc}$ because it makes the loss term smaller than what it would be if ERE was constant or decreased. So, overall, the $V_{oc}$ shows a slightly increasing trend with decreasing temperature.

$$V_{\text{oc}} = \frac{E_g}{q} - \Sigma V_{loss} + \frac{kT}{q}\ln(ERE) \qquad (1)$$

In contrast, the reduction in FF is likely due to limitations in charge carrier transport and extraction at low temperatures, which also result in increased series resistance.[32,40] Similar trends in device parameters at low temperatures have been reported in previous studies on mixed-cation perovskites, where charge transport and extraction limitations were linked to FF degradation and $R_s$ growth.[39] The PCE initially increases from 22.08 % at 300 K to 22.91 % at 280 K, before declining sharply to 9.77 % at 160 K. This initial improvement is attributed to the increase in $V_{oc}$ with a slight decrease in temperature, while the later reduction in PCE is dominated by the strong decline in FF at lower temperatures. Additionally, the light *J-V* curves begin to exhibit an S-shaped form below 240 K, indicating a nonlinear resistance to the flow of charge through the device. Previous studies have reported that such S-shaped light *J-V* characteristics result from hindered charge extraction in MAPbI$_3$ perovskite solar cells at low temperatures, consisting with the formation of an energy barrier or a schottky diode at one or both interfaces.[40] Therefore, the low-temperature electrical results in **Figure 3**a,b are consistent with the EL measurements where charge must be injected into the active layer from the contacts. However, since PL is achieved via direct light excitation of the active layer, the contact injection limitations are not critical and the PL ERE measurements probe the intrinsic properties of the perovskite layer.

Injection Current-Dependent EL Imaging

To further investigate the EL behavior in perovskite devices, we performed injection current-dependent EL ERE mapping across the entire cell area ($\approx$ 7.2 mm$^2$). These measurements were conducted at 300 K and 200 K under injection currents of 0.3 mA, 1 mA, and 4 mA, as shown in **Figure 4**a. At 300 K and 0.3 mA, the EL ERE map appears relatively uniform, except for a few pinholes and edge defects, indicated by blue/yellow regions on the device. As the injection current is increased to 1 mA and then 4 mA, the overall EL ERE across the device systematically increases, and spatial variations become more pronounced. At 200 K, regions with suppressed charge injection are clearly visible in the ERE map. We note that each image has its own unique colorbar scale, and hence comparing pixels or regions across temperatures or injection currents requires careful attention to the actual local ERE value, not the color of the pixels. For example, comparing the 4 mA images at 300 K and 200 K, one can find several small red regions in the 200 K image that have ERE values greater than their value at 300 K, even though a large part of the device has been degraded into dark blue regions at 200 K. This observation indicates that contact resistance or the energy barrier to charge injection is not uniform across the device and has a local characteristic.

**Figures S5** and **S6** (Supporting Information) show the temperature dependence of the ERE for several local areas as labeled in **Figure S5**a-d (Supporting Information). These regions labeled P1 through P5 show that consistent with the mean EL ERE vs. T measurements, local EREs initially increase as T is lowered but most spots start falling below 240 K, except for a few regions where the decrease is pushed deeper into the low-T regime (e.g., P2). These local observations indicate that if a local contact region is good, the local EL ERE can continue to increase with decreasing temperatures much like in the PL ERE measurements. However, even the most efficient spots eventually start to show a decrease in the ERE as T is lowered, because charge injection over an energy barrier is temperature-dependent, and even a low local energy barrier can become a problem at a low enough temperature. Therefore, even though the perovskite material itself radiatively becomes more efficient at low temperatures due to a deactivation of nonradiative defect channels, the charge injection problem in EL measurements will still prevent the EL ERE from increasing as T is lowered.

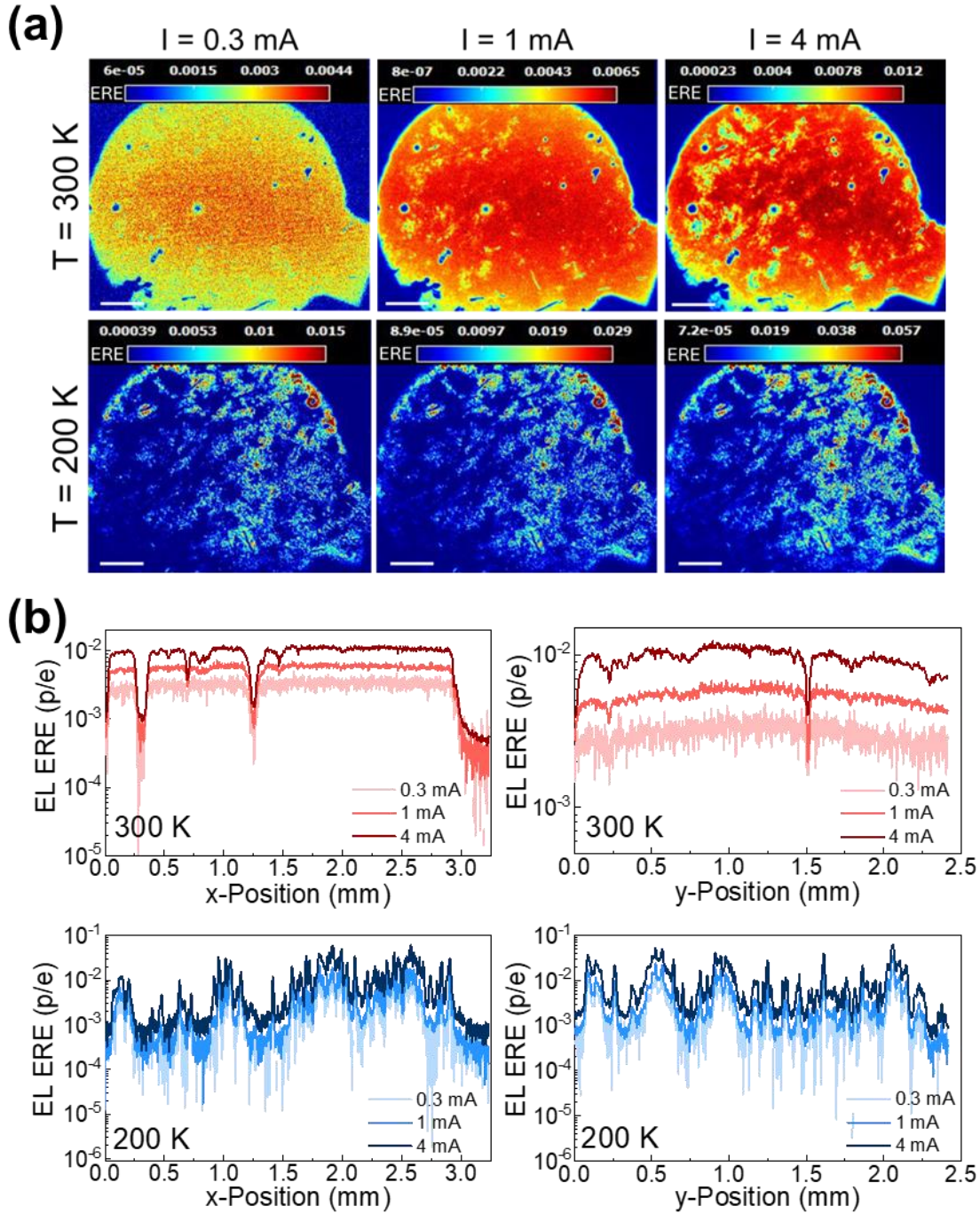

**Figure 4.** (a) Injection current dependence of EL ERE imaging over an entire device, and (b) EL ERE line scans at 300 K and 200 K. Scalebar is 500 μm.

**Figure 4**b shows the x- and y-direction line scans of the EL images (scan positions in **Figure S7**, Supporting Information), demonstrating that while the 300 K scans show a mostly uniform ERE behavior at relatively high values (ERE ≈ $10^{-2}$), the lower temperature scans reveal significant micro-fluctuations and variability in ERE values. These results confirm that charge injection issues have a local nature and are not felt homogeneously across the entire device area. **Figure S8** (Supporting Information) displays the average EL ERE versus injection current density at various temperatures. At 300 K, the EL ERE is approximately $10^{-3}$ under low current densities ($J_{inj} \approx 1$ mA/cm$^2$), increasing substantially with current, reaching around $10^{-2}$ at $J_{inj} >$ 10 mA/cm². This current density dependence of ERE in perovskite PV cells has been extensively discussed in our previous work[41] and is attributed to the competition between radiative and non-radiative recombination channels. As current increases, radiative recombination becomes more dominant in the active layer, resulting in higher ERE values. At lower temperatures, while the current dependence of ERE persists, the behavior becomes more complex due to the nonlinear contact resistance phenomena discussed here.

3. Conclusion

In this study, we systematically investigated the temperature dependence of the luminescence response of inverted RbCsMAFA-based PSCs via hyperspectral EL and PL imaging. These measurements were further complemented with temperature-dependent dark and light *J-V* characterizations. Our results reveal complex correlations between charge carrier recombination dynamics and local charge injection effects, leading to spatial inhomogeneity in the luminescence efficiency across the device and distinctive EL and PL temperature-dependent ERE plots and images. Both EL and PL ERE improve as the temperature is lowered from 300 K to 240 K due to the suppression of non-radiative recombination. However, further cooling below 240 K results in a decline in EL ERE, which is attributed to suppressed charge carrier injection and extraction via the perovskite/contact layer interfaces. Although recombination within the perovskite layer remains efficient, as evidenced by good PL ERE at all temperatures, injection-related losses dominate EL behavior below 240 K. Overall, our results highlight the importance of understanding temperature triggered charge injection and transport restriction in perovskite devices. We find that temperature-dependent EL imaging, even without an absolute scale, can serve as a useful tool for understanding and mapping local charge injection issues within the device stack. These findings are critical to providing a deeper understanding of the low-temperature performance of mixed cation halide perovskite solar cells, creating a pathway for the potential use of these materials in space applications.

## 4. Experimental Section/Methods

*4.1. Fabrication of Perovskite Solar Cells*:

Materials: Methylammonium bromide (MABr; 99.99 %) and formamidinium iodide (FAI; 99.99 %) were procured from Greatcell Solar Materials. Rubidium iodide (RbI; 99.9 %) and semicarbazide hydrochloride (SECI, 99 %) were sourced from Sigma-Aldrich, while cesium iodide (CsI; 99.999 %) was purchased from Sigma-Aldrich. Lead bromide (PbBr$_2$; 99.99 %), lead iodide (PbI$_2$; 99.99 %), 1,3-diaminopropane dihydroiodide (PDAI$_2$; 98 %), and [2-(3,6-dimethoxy-9H-carbazol-9-yl)ethyl]phosphonic acid (MeO-2PACz; 98 %) were ordered from TCI America. All high-purity and anhydrous solvents, including dimethyl sulfoxide (DMSO; 99.9 %), N, N-dimethylformamide (DMF; 99.8 %), chlorobenzene (99.8 %), 2-propanol (99.5 %), and ethanol (99.5 %), were obtained from Sigma-Aldrich.

Perovskite Precursor Preparation: The perovskite precursor solution was prepared in a nitrogen-filled glovebox (< 0.5 ppm O$_2$ and H$_2$O) to avoid ambient air exposure. The targeted perovskite composition was *FA$_{0.85}$MA$_{0.05}$Cs$_{0.05}$Rb$_{0.05}$Pb(I$_{0.95}$Br$_{0.05}$)$_3$* [CsRbFAMA], with a total concentration of 1.5 M. To formulate the precursor solution, RbI (15.9 mg), CsI (19.5 mg), MABr (8.4 mg), PbBr$_2$ (27.5 mg), FAI (219.3 mg), PbI$_2$ (656.9 mg), and SECI (1.5 mg) as an additive, were accurately weighed and subsequently dissolved in 1 mL of a DMF: DMSO solvent mixture (4:1 volume ratio). The precursor solution was thoroughly mixed using a magnetic stirrer for 1 hour and then filtered using a 0.22 µm polytetrafluoroethylene (PTFE) syringe filter before use.

Device Fabrication: Patterned Fluorine-doped tin oxide (FTO) substrates (TEC 15, 24.9 mm by 24.9 mm) were sequentially underwent an ultrasonic bath in Micro-90 liquid cleaning solution, deionized water (twice), acetone, and 2-propanol, each for 30 min, and then dried with compressed air. Before spin-coating, the substrates were treated with ultraviolet–ozone for 30 min for better surface wettability. All steps were performed inside a nitrogen-filled glovebox. 100 µL MeO-2PACz (0.75 mg/mL in ethanol) hole-transporting layer was spin-coated onto the cleaned FTO substrates at a spin rate of 3000 rpm (1 rpm = 0.105 rad/s) for 30 s, followed by thermal annealing at 100 °C for 10 min. After cooling to ambient temperature, 70 µL of the filtered CsRbFAMA perovskite precursor solution was deposited onto the hole transport layer via a two-step spin-coating process: initially at 1000 rpm for 10 s, and then at 4000 rpm for 40 s. At 30 s into the second step, 150 µL chlorobenzene (antisolvent) was dropped onto the spinning substrates (about 1 cm distance), and then annealed at 100 °C for 10 min. After cooling, the perovskite layers were passivated by dynamically spin-coating 100 µL of PDAI$_2$ (0.3 mg/mL in 2-propanol) at a speed of 4500 rpm for 30 s, and then finally annealed for 5 min. Thereafter, the substrates were transferred to the thermal evaporation system for $C_{60}$ (25 nm), BCP (5 nm), and Ag (100 nm) deposition at rates of 0.15, 0.05, and 0.50 Å/s (1 Å = 100 pm), respectively.

*4.2. Electroluminescence and photoluminescence imaging measurements:*

EL and PL measurements of the perovskite solar cells were performed using a Photon ETC Grand EOS hyperspectral wide-field imaging system. EL was performed at various applied currents using a source meter at different temperatures ranging from 300 K to 150 K. PL was conducted by light excitation using a 532-nm laser over the entire field of view, typically several mm in extent. The laser intensity was kept under 10 mW/cm$^2$ to minimize heating or damage to the material. The spectral range of the hyperspectral measurements was from 700 nm to 900 nm. The spectral resolution is about 2 nm, and all images were taken through a 5× objective lens. For temperature dependent measurements, the perovskite cell was placed inside an optical cryostat and was kept under vacuum at $10^{-6}$ Torr. The temperature of the device was controlled with a Lake Shore Model 335 temperature controller. Detailed information about the calibration of the EL system to obtain absolute photon flux rates is given in our previous work.[42] The relative uncertainties of our photon flux measurements are estimated to be around ± 15 %.

*4.3. Electrical measurements:* The dark *J-V* measurements were carried out as a function of temperature typically from 300 K to 150 K using the steady state *J-V* approach[43] with a Keithley model 2601 source-measure unit. Light *J-V* characterization was conducted under a xenon solar simulator under the standard reporting conditions (i.e., 1000 W/m², the air mass 1.5 global, and ≈ 25 °C). A NIST calibrated silicon cell was used as the reference cell for light *J-V* characterization, and the *J-V* measurements were spectral-mismatch corrected according to established procedures. The spectral mismatch parameter is typically around 0.94 at room temperature. An uncertainty analysis of the *J-V* curve parameters at room temperature using our measurement system reveals a relative expanded uncertainty of 2.14 % for the $I_{sc}$, 0.6 % for the $V_{oc}$, 3.32 % for the FF and 3.2 % for the PCE.

*4.4. EQE measurements:* EQE measurement was taken in the range of 280 nm to 1000 nm using a monochromator-based differential spectral responsivity system. The details of the system operation have been provided in our previous work.[44]

Acknowledgements

H.Y-C. acknowledges the generous support of the Professional Research Experience Program (PREP) under the award number PREP0002351. This material is based on research sponsored by the Air Force Research Laboratory under agreement numbers FA9453-19-C-1002 and FA9453-24-C-X004.

NIST disclaimer: Certain equipment, instruments, software, or materials are identified in this paper in order to specify the experimental procedure adequately. Such identification is not intended to imply recommendation or endorsement of any product or service by NIST, nor is it intended to imply that the materials or equipment identified are necessarily the best available for the purpose.

DoD disclaimer: The U.S. Government is authorized to reproduce and distribute reprints for Governmental purposes notwithstanding any copyright notation thereon. The views expressed are those of the authors and do not reflect the official guidance or position of the United States Government, the Department of Defense or of the United States Air Force. The appearance of external hyperlinks does not constitute endorsement by the United States Department of Defense (DoD) of the linked websites, or the information, products, or services contained therein. The DoD does not exercise any editorial, security, or other control over the information you may find at these locations.